\documentclass[conference,letterpaper]{IEEEtran}

\usepackage[T1]{fontenc}
\usepackage[utf8]{inputenc}
\usepackage{amsmath,amssymb,amsfonts,bm}
\usepackage{mathtools}
\usepackage{graphicx}
\usepackage{booktabs}
\usepackage{multirow}
\usepackage{xcolor}
\usepackage{cite}
\usepackage{url}
\usepackage{capt-of}
\usepackage{algorithm}
\usepackage{algpseudocode}
\usepackage{siunitx}
\usepackage{ifthen}
\usepackage[letterpaper,left=1.02in,right=1.02in,top=0.75in,bottom=1.05in]{geometry}

\graphicspath{{./}}
\DeclareMathOperator*{\argmin}{arg\,min}

\newcommand{\vect}[1]{\boldsymbol{#1}}
\newcommand{\mat}[1]{\mathbf{#1}}
\newcommand{\set}[1]{\left\{#1\right\}}

\newcommand{\C}{\mathbb{C}}
\newcommand{\Eset}{\mathcal{E}}
\newcommand{\Yset}{\mathcal{Y}}
\newcommand{\Mpri}{\mathcal{M}_{\mathrm{pri}}}
\newcommand{\Knbrset}{\mathcal{K}_{\mathrm{nbr}}}
\newcommand{\paperfigwidth}{\columnwidth}

\title{A Reduced Greedy Minimum Distance Phase Shift Design for Quantized RIS-Assisted\\ Spatial Modulation}

\author{\IEEEauthorblockN{Alexander Kelly and Mark F. Flanagan}
\IEEEauthorblockA{School of Electrical and Electronic Engineering, University College Dublin, Belfield, Dublin 4, Ireland\\
Email: alexander.kelly1@ucdconnect.ie, mark.flanagan@ieee.org}}

\begin{document}
\maketitle

\begin{abstract}
Reconfigurable intelligent surfaces (RISs) have emerged as a key enabler for next-generation wireless networks, offering control over the propagation environment through nearly passive reflecting elements. When combined with spatial modulation (SM), an RIS shapes a finite set of noise-free signal points at the receiver, and the error rate performance under maximum-likelihood (ML) detection is governed by the smallest pairwise Euclidean distance among these points. In this paper, we propose a low-complexity RIS phase shift design algorithm for RIS-assisted SM under practical discrete phase shift control. The proposed method targets the minimum Euclidean distance directly through a reduced greedy search over the discrete set of RIS phase shifts, in contrast to existing approaches that align RIS phase shifts to maximize the effective channel gain and therefore do not directly address the dominant error events. The search effort is controlled by a small set of integer parameters, providing a tunable trade-off between bit error rate (BER) performance and computational cost. We evaluate the proposed method against continuous and quantized phase-alignment baselines, including a complexity-matched comparison against the most directly comparable benchmark. Numerical results show that the proposed method achieves a significant SNR gain over quantized phase alignment under the same discrete hardware model, and this advantage persists under imperfect channel state information.
\end{abstract}


\section{Introduction}
Reconfigurable intelligent surfaces (RISs) represent a promising enabling technology for next-generation wireless networks. An RIS is a planar array of low-cost, nearly passive reflecting elements, each capable of applying an independently controlled phase shift to an incident electromagnetic wave~\cite{Basar2019RIS,DiRenzo2020SRE}. By coordinating these phase shifts, the RIS can shape the propagation environment, providing coverage in non-line-of-sight regions, mitigating fading, and improving received signal strength without active radio-frequency hardware~\cite{Basar2019RIS}. Because the elements are nearly passive, RIS technology offers these gains at low energy cost relative to fully active arrays, which is one of its main attractions for energy-constrained deployments.

Spatial modulation (SM) is a multi-antenna signaling technique that also targets hardware and energy efficiency~\cite{Mesleh2008SM,DiRenzo2011SSK,Sugiura2011GSTSK,Basar2017IM}. Rather than transmitting additional information through IQ symbols, these schemes convey additional information through the index of an active transmit antenna or antenna subset. As a result, far fewer active radio-frequency chains are required than in a conventional MIMO system with the same throughput, while inter-channel interference is also avoided. 

The combination of these two technologies has been studied increasingly in recent years~\cite{Canbilen2020RISSSK,Li2021RISSSK,Yue2023TFSSK}. In an RIS-assisted single-antenna link, the RIS serves to enhance the strength of the (single) effective channel. In contrast, in an RIS-assisted SM link, the transmitter switches between different antennas (or antenna subsets) according to the data, so the RIS phase shift configuration influences a whole family of effective channels in parallel. The geometry of the resulting set of noise-free signal points at the receiver, rather than the gain of any one channel, is the key metric affecting performance.

This distinction is important, because the optimum receiver in such systems is the maximum-likelihood (ML) detector, which compares the noisy received sample to a finite set of noise-free signal points and selects the closest match. Under ML detection, dominant error events occur between the closest pairs of points in this set~\cite{DiRenzo2011SSK,Sugiura2011GSTSK}. The relevant performance metric is therefore the global minimum Euclidean distance between these points, not the magnitudes of the points themselves. Despite this, most RIS design methods proposed in the literature for SM are gain-oriented, i.e., the RIS phase shifts are chosen to maximize the received signal strength~\cite{Canbilen2020RISSSK,Yue2023TFSSK}. While simple and effective for the case of a single-antenna transmitter, gain-oriented designs do not directly address the minimum Euclidean distance and may leave the dominant error events largely unchanged. A more recent line of work has begun to consider distance-based design rules for RIS-assisted SSK~\cite{Li2021RISSSK,Zhu2024DiscreteRISSSK}; however, these methods are not directly applicable to the scenario of SM.

A second practical issue is that, due to the nature of the underlying components, physical RIS hardware does not provide continuous phase shifts. Instead, each element can apply only one of a finite number of phase shifts~\cite{Sang2024QuantisedPhase,Wu2020Discrete,Xu2021QuantLevels}. Phase shift design under this constraint is a discrete combinatorial optimization problem whose search space grows exponentially with the number of RIS elements and the phase resolution, making exhaustive search infeasible for any practical RIS size. A range of reduced-complexity approaches has been considered, including local-search heuristics, alternating optimization, and continuous-then-quantized relaxations~\cite{Wu2020Discrete,Sang2024QuantisedPhase,Pradhan2024Probabilistic}.

Two gaps remain in the literature. First, in an RIS-assisted SM setting with ML detection, no proposed phase shift design rule directly targets the minimum Euclidean distance under a discrete phase alphabet. Second, where reduced-complexity heuristics are proposed, the resulting trade-off between BER performance and the computational budget allocated to the search is rarely
exposed explicitly. 

In this paper, we propose a phase shift design method for RIS-assisted SM that targets the global minimum Euclidean distance of the ML receive-point set under discrete RIS phase shift control. The main contributions of this paper are summarized as follows.
\begin{itemize}
\item We formulate a discrete max--min distance based phase shift design problem for RIS-assisted SM, in which the RIS phase shift configuration is selected so as to maximize the global minimum Euclidean distance under a finite phase shift alphabet. 
\item We propose a reduced greedy search algorithm that approximately solves the formulated problem, exploiting three key ideas: concentrating updates on the receive points with the smallest nearest-neighbor distance, considering only the RIS elements with the largest cascaded channel gains for each such point, and replacing the global distance comparison with a local one over a small set of nearest neighbors. Four integer parameters control the search space size, leading to a tunable BER--complexity trade-off.
\item We evaluate the performance of the proposed scheme against that of the most directly relevant benchmark from the literature, under both continuous and quantized phase alignment variants, providing in addition a comparison under an overall complexity constraint. Our results show that the proposed method achieves a significant SNR gain over the benchmark scheme with quantized phase alignment, and that the BER advantage persists under the constraint of matched receiver complexity. 
\item We evaluate the robustness of the proposed method under imperfect channel state information (CSI), in which the RIS is configured using the estimated channels and the ML detector operates using an estimated receive point set. Across the CSI error levels tested, the proposed method continues to outperform the gain-oriented baseline schemes.
\end{itemize}

\section{System Model}\label{sec:system}
An SM system is considered in which a transmitter equipped with $N_t$ antennas communicates with a single-antenna receiver\footnote{For ease of exposition, a single-antenna receiver is considered in this paper. However, it should be noted that the proposed method can be straightforwardly generalized to the case of a multi-antenna receiver. In this case, the Euclidean distance to be minimized is in a higher-dimensional space, but other details of the proposed algorithm are the same.} via an RIS with $N$ reflecting elements. The transmitter-to-RIS channel is denoted by $\mat{H} \in \C^{N_t \times N}$, with $H_{t,n}$ denoting its $(t,n)$-entry, and the RIS-to-receiver channel is denoted by $\vect{g} \in \C^{N}$, with $g_n$ denoting its $n$-th entry. The entries of $\mat{H}$ and $\vect{g}$ are i.i.d. and distributed according to\ $\mathcal{CN}(0,1)$, corresponding to a Rayleigh fading model. The direct transmitter-to-receiver path is assumed to be blocked or sufficiently weak to be neglected, which models the deployment scenario for which an RIS is most useful in practice~\cite{DiRenzo2020SRE}.

Each RIS element applies one phase shift drawn from a finite alphabet\footnote{The proposed method does not rely on the structure of $\Eset$ and can be applied to an arbitrary phase shift set, including phase shift sets with non-unit amplitudes.}
\begin{equation}
\Eset = \set{q_\ell = e^{j 2\pi \ell / L}\,:\, \ell = 0,1,\ldots,L-1},
\end{equation}
of size $L$, where $L = 2^b$ for an RIS supporting $b$-bit phase shift control.

According to the SM principle, information is conveyed through the index of the active transmit antenna subset. Let $\mathcal{T} = \set{\mathcal{T}_1,\ldots,\mathcal{T}_M}$ denote the set of admissible transmit antenna subsets, where each $\mathcal{T}_m \subseteq \set{1,\ldots,N_t}$ contains the indices of the antennas activated for the $m$-th subset. The number of active antennas in subset $\mathcal{T}_m$ is denoted by $N_a(m) = |\mathcal{T}_m|$. The cardinality $M$ is usually chosen to be a power of two, so that an integer number of bits is mapped to a subset index. As an illustrative example, with $N_t=4$ and $M=8$, one may choose the four singletons $\set{1}, \set{2}, \set{3}, \set{4}$ and the four pairs $\set{1,2}, \set{1,3}, \set{1,4}, \set{2,3}$, giving eight admissible subsets that involve only one or two antenna activations. At the transmitter, the incoming bits select one subset index $m$, and only the antennas in $\mathcal{T}_m$ are activated.

For a given transmission, let $m$ denote the selected subset index. Denote by $\bar{h}_{m,n}$ the equivalent transmitter-to-RIS channel coefficient seen at RIS element $n$, formed by combining the contributions from the active antennas with energy normalization:
\begin{equation}
\bar{h}_{m,n} = \frac{1}{\sqrt{N_a(m)}} \sum_{t \in \mathcal{T}_m} H_{t,n}.
\end{equation}
The normalization by $\sqrt{N_a(m)}$ ensures that the transmitted energy does not depend on the subset size. Combining this equivalent channel coefficient with the corresponding RIS-to-receiver channel coefficient gives the cascaded channel coefficient
\begin{equation}
a_{m,n} = \bar{h}_{m,n}\, g_n.
\end{equation}
The squared magnitude $\alpha_{m,n} = |a_{m,n}|^2$ is referred to as the \emph{cascaded channel gain} via element $n$ for the $m$-th subset. Collecting all coefficients yields $\mat{A} = [a_{m,n}] \in \C^{M \times N}$.

If RIS element $n$ is configured with phase shift index $\omega_{m,n} \in \set{0,1,\ldots,L-1}$ when subset $m$ is transmitted, then the corresponding noise-free received signal point is
\begin{equation}\label{eq:ym_def}
y_m = \sum_{n=1}^{N} a_{m,n}\, q_{\omega_{m,n}},
\end{equation}
and the receive-point set, i.e., the set of noise-free signal points observed at the receiver taking into account the $M$ admissible subsets, is
\begin{equation}\label{eq:Yset_def}
\Yset = \set{y_1, y_2, \ldots, y_M}.
\end{equation}
The RIS phase shift index matrix $\mat{\Omega} = [\omega_{m,n}] \in \set{0,\ldots,L-1}^{M \times N}$, whose $(m,n)$-entry is $\omega_{m,n}$, collects all global phase shift choices and parameterizes $\Yset$.

\subsection{Maximum-Likelihood Detection}
After transmission over the RIS-aided channel, the received sample is
\begin{equation}
r = y_{m} + w,
\end{equation}
where $m$ is the chosen transmitted subset index and $w \sim \mathcal{CN}(0,\sigma_w^2)$ represents complex additive white Gaussian noise. Equiprobable subsets are assumed. The ML detector selects the index whose noise-free receive point lies closest in Euclidean distance to $r$:
\begin{equation}\label{eq:ML}
\hat{m} = \argmin_{\tilde{m} \in \set{1,\ldots,M}} |r - y_{\tilde{m}}|^2,
\end{equation}
with $\hat{m}$ denoting the receiver's decision.

The detection rule in~\eqref{eq:ML} makes the relative locations of the elements of $\Yset$ the main determinant of error performance. For each receive point $y_m$, the \emph{$m$-local minimum distance}
\begin{equation}\label{eq:dm}
d_m = \min_{i \neq m}\, |y_m - y_i|^2
\end{equation}
captures the closest nearest neighbor of $y_m$ in $\Yset$, and the global minimum distance is given by
\begin{equation}\label{eq:dmin}
d_{\min} = \min_m d_m = \min_{i \neq j}\, |y_i - y_j|^2.
\end{equation}
For system design targeting low BER, $d_{\min}$ is a natural design choice for RIS phase shift optimization, as it directly captures the effect of the dominant error events.

\section{Reduced Greedy Minimum-Distance Phase Shift Design}\label{sec:gmd}

\subsection{Design Problem}
The objective is to select the RIS phase shift configuration in order to maximize the global minimum Euclidean distance of the receive-point set $\Yset$:
\begin{equation}\label{eq:GMD_problem}
\max_{\mat{\Omega} \in \set{0,\ldots,L-1}^{M \times N}}\; \min_{i \neq j}\, |y_i - y_j|^2,
\end{equation}
where $y_i$ and $y_j$ are functions of $\mat{\Omega}$ through~\eqref{eq:ym_def}. A related max--min distance objective has been used for design of RIS-assisted SSK with a single active transmit antenna~\cite{Li2021RISSSK,Zhu2024DiscreteRISSSK}; the formulation here extends this objective to the case of a multi-antenna transmitter as well as to a discrete phase shift alphabet. The problem in~\eqref{eq:GMD_problem} is a discrete combinatorial optimization problem with search space size $L^{MN}$; for example, for the system parameters considered in Section~\ref{sec:results} ($M=8$, $N=32$, $L=4$), the search space contains $4^{8\times32}\approx 1.4\times10^{154}$ configurations, making exhaustive search infeasible. A reduced greedy search procedure that approximates~\eqref{eq:GMD_problem} is described in the following subsection.

\subsection{Reduced Greedy Search}
The proposed method finds a suboptimal solution to~\eqref{eq:GMD_problem} using a coordinate-wise greedy procedure that updates one entry of $\mat{\Omega}$ at a time, starting from an initial configuration $\mat{\Omega}^{(0)}$. The complete procedure is given in Algorithm~\ref{alg:gmd_main}. At the start (lines 3--4), the receive points $y_m$ are initialized from $\mat{\Omega}^{(0)}$ and the cascaded channel gains $\alpha_{m,n}$ are precomputed. In each pass $p$ (lines 5--21), the algorithm identifies the receive points and RIS elements on which to concentrate updates, then calls the subroutine \textsc{PhaseUpdate} (detailed in Algorithm~\ref{alg:phase_update}) for each selected $(m,n)$ pair. If the subroutine returns a new phase shift index, the update is accepted and $\mat{\Omega}$, $y_m$, and $d_m$ are revised (line 15). The pass terminates early if no update was accepted (line 20). Up to $P$ passes are performed.

Three key observations are exploited to concentrate the search where it is most likely to improve $d_{\min}$:

\emph{Receive points worth updating (lines 6--7 of Algorithm~\ref{alg:gmd_main}).} The value of $d_{\min}$ is determined by the closest pair in $\Yset$. Receive points whose $m$-local minimum distance $d_m$ is already large contribute little to $d_{\min}$, whereas those with the smallest $d_m$ are exactly the receive points that determine $d_{\min}$. The procedure therefore selects only $K_{\mathrm{p}}$ receive points with the smallest $d_m$ as the \emph{priority set} $\Mpri$ in each pass $p$.

\emph{RIS elements worth testing (line 11).} For a fixed receive point $y_m$, changing the phase shift index of RIS element $n$ from $\omega_{m,n}$ to $\ell$ displaces $y_m$ by $a_{m,n}(q_\ell - q_{\omega_{m,n}})$, whose magnitude is bounded by $2|a_{m,n}|$. Elements with the largest cascaded channel gains $\alpha_{m,n} = |a_{m,n}|^2$ therefore offer the largest possible displacements of $y_m$. The procedure selects only the $K_{\mathrm{e}}$ elements with the largest $\alpha_{m,n}$ for each $m \in \Mpri$, forming the set $\mathcal{N}_m$.

\emph{Local distance checks (lines 9--14 of Algorithm~\ref{alg:phase_update}).} When evaluating a tentative phase shift change for $y_m$, the global minimum distance is determined either by pairs not involving $y_m$ (precomputed as $d_{\setminus m}$ in line 10 of Algorithm~\ref{alg:gmd_main}) or by the distances between the updated location of $y_m$ and the other receive points. Rather than comparing against all $M-1$ other points, the proposed method checks only the $K_{\mathrm{l}}$ current nearest neighbors of $y_m$, since these are the points most likely to determine the minimum distance after a small coordinate update. Setting $K_{\mathrm{l}} = M-1$ recovers the full comparison.

\begin{algorithm}[t]
\caption{Reduced Greedy Phase Shift Design for Minimum Distance Maximization}
\label{alg:gmd_main}
\footnotesize
\begin{algorithmic}[1]
\State \textbf{Input:} coefficient matrix $\mat{A}=[a_{m,n}]$, alphabet $\Eset$, initial $\mat{\Omega}^{(0)}$, integers $P$, $K_{\mathrm{p}}$, $K_{\mathrm{e}}$, $K_{\mathrm{l}}$
\State \textbf{Output:} phase shift index matrix $\mat{\Omega}^\star$, receive-point set $\Yset^\star$
\State $y_m \gets \sum_{n} a_{m,n} q_{\omega_{m,n}}$ for all $m$ \Comment{initialize receive points}
\State $\alpha_{m,n} \gets |a_{m,n}|^2$ for all $(m,n)$ \Comment{precompute cascaded gains}
\For{$p = 1$ to $P$}
    \State $d_m \gets \min_{k\neq m}|y_m - y_k|^2$ for all $m$
    \State $\Mpri \gets$ indices of the $K_{\mathrm{p}}$ smallest $\{d_m\}$ \Comment{priority set}
    \State $\texttt{changed} \gets \texttt{False}$
    \ForAll{$m \in \Mpri$}
        \State $d_{\setminus m} \gets \min_{i<j,\, i,j\neq m} |y_i - y_j|^2$
        \State $\mathcal{N}_m \gets$ indices of the $K_{\mathrm{e}}$ largest $\{\alpha_{m,n}\}_{n=1}^N$
        \ForAll{$n \in \mathcal{N}_m$}
            \State \makebox[0pt][l]{$(\omega^\star, y_m^\star, d_m^\star, g^\star) \gets$}
            \Statex \hspace{\dimexpr\algorithmicindent+3em\relax}$\Call{PhaseUpdate}{m,n,\mat{A},\Yset,\mat{\Omega},d_{\setminus m},d_m,K_{\mathrm{l}}}$
            \If{$\omega^\star \neq \omega_{m,n}$} \Comment{accept if improved}
                \State $\omega_{m,n}\!\gets\!\omega^\star$,\ $y_m\!\gets\!y_m^\star$,\ $d_m\!\gets\!d_m^\star$
                \State $\texttt{changed} \gets \texttt{True}$
            \EndIf
        \EndFor
    \EndFor
    \If{$\texttt{changed} = \texttt{False}$} \textbf{break} \EndIf \Comment{early stopping}
\EndFor
\State \Return $(\mat{\Omega},\Yset)$
\end{algorithmic}
\end{algorithm}

\subsection{Single-Coordinate Update Subroutine}
Algorithm~\ref{alg:phase_update} is the subroutine \textsc{PhaseUpdate} called at line 13 of Algorithm~\ref{alg:gmd_main}. For a given pair $(m,n)$, it exhaustively tests all $L$ phase shift values for RIS element $n$ and returns the one that best improves $d_{\min}$, keeping all other entries of $\mat{\Omega}$ fixed. The $K_{\mathrm{l}}$ receive points nearest to $y_m$ are identified in lines 3--4 and used as a reduced set of nearest neighbors. The current contribution of element $n$ is removed from $y_m$ (line 4), and each candidate phase shift value $q_\ell$ is tested in turn (lines 7--11). For each candidate, the updated receive point $\tilde{y}_m$ and its m-local distance $\tilde{d}_m$ are computed (lines 7--8), and the resulting global minimum $\tilde{g}=\min(d_{\setminus m},\tilde{d}_m)$ is compared to the current minimum value $g^\star$ (lines 9--12). A phase shift change is accepted if and only if $\tilde{g}>g^\star$, with a tie-break on $\tilde{d}_m$ used in the case $\tilde{g}=g^\star=0$ (representing the collision of two receive points). Since only $y_m$ changes during any one call, the value $d_{\setminus m}$ passed in as input remains valid throughout, avoiding a full recomputation of all pairwise distances.

\begin{algorithm}[t]
\caption{Subroutine \textsc{PhaseUpdate}: Single-Coordinate Phase Shift Optimization}
\label{alg:phase_update}
\footnotesize
\begin{algorithmic}[1]
\State \textbf{Input:} indices $m$, $n$, $\mat{A}$, $\Yset$, $\mat{\Omega}$, minimum distance excluding $y_m$, $d_{\setminus m}$, current m-local distance of $y_m$, $d_m$, number of nearest neighbors $K_{\mathrm{l}}$
\State \textbf{Output:} best phase shift index $\omega^\star$, updated receive point $y_m^\star$, updated distance $d_m^\star$, updated global minimum $g^\star$
\State $\Knbrset(m) \gets$ indices of the $K_{\mathrm{l}}$ nearest receive points to $y_m$
\State $y^{-} \gets y_m - a_{m,n}\, q_{\omega_{m,n}}$ \Comment{remove current contribution}
\State $(\omega^\star, y_m^\star, d_m^\star, g^\star) \gets (\omega_{m,n},\, y_m,\, d_m,\, \min(d_{\setminus m},\,d_m))$
\For{$\ell = 0$ to $L-1$} \Comment{test all phase shift values}
    \State $\tilde y_m \gets y^{-} + a_{m,n}\, q_\ell$
    \State $\tilde d_m \gets \min_{k \in \Knbrset(m)} |\tilde y_m - y_k|^2$
    \State $\tilde g \gets \min(d_{\setminus m},\, \tilde d_m)$
    \If{$\tilde g > g^\star$ \textbf{or} ($\tilde g = g^\star = 0$ \textbf{and} $\tilde d_m > d_m^\star$)}
        \State $(\omega^\star, y_m^\star, d_m^\star, g^\star) \gets (\ell,\, \tilde y_m,\, \tilde d_m,\, \tilde g)$
    \EndIf
\EndFor
\State \Return $(\omega^\star, y_m^\star, d_m^\star, g^\star)$
\end{algorithmic}
\end{algorithm}

\subsection{Computational Complexity}
At most $P$ passes are performed; in each pass, $K_{\mathrm{p}}$ receive points are selected; for each, $K_{\mathrm{e}}$ elements are tested; and at each candidate coordinate, $L$ phase shift values are evaluated against $K_{\mathrm{l}}$ nearest neighbors. Thus, the complexity of the overall reduced greedy search algorithm per channel realization is $\mathcal{O}(P K_{\mathrm{p}} K_{\mathrm{e}} K_{\mathrm{l}})$. 

\section{Numerical Results}\label{sec:results}

\subsection{Simulation Setting}
We use the following simulation parameters unless otherwise stated. The transmitter is equipped with $N_t=4$ antennas and $M=8$ admissible transmit subsets are used, comprising the four singletons $\{1\}$, $\{2\}$, $\{3\}$, $\{4\}$ and the four pairs $\{1,2\}$, $\{1,3\}$, $\{1,4\}$, $\{2,3\}$. The RIS comprises $N=32$ elements with $b=2$-bit phase shift control, so that $L=4$. The BER is averaged over at least 200 independent channel realizations.

As the principal benchmark, we adopt two variants of the RIS-assisted SM optimization method of~\cite{Yue2023TFSSK}. The first benchmark, which we term \textit{continuous phase alignment} (CPA) is the scheme originally proposed in~\cite{Yue2023TFSSK}. Under CPA, the phase shifts are chosen so that every cascaded channel coefficient adds coherently at the receiver; this represents an idealized, hardware-unconstrained reference. The second benchmark, here termed as \textit{quantized phase alignment} (QPA) represents a straightforward modification of the method of ~\cite{Yue2023TFSSK} for the case of quantized phase shifts. This method applies the same coherent-combining rule but quantizes each phase shift to the nearest entry of E, and therefore operates under the same discrete hardware model as the proposed method. Consequently, any BER difference between the method of~\cite{Yue2023TFSSK} with QPA and the proposed method may be attributed to the RIS phase shift design approach rather than to the hardware model.

Unless otherwise stated, the proposed method uses the reference setting\footnote{A gain-ranked variant of the
proposed method, in which the initial phase shift index matrix $\mat{\Omega}^{(0)}$ is ordered by mean cascaded channel gain, was also investigated. It was observed that this provided no discernible improvement, so the fixed-mapping variant is used throughout.} $(P,K_{\mathrm{p}},K_{\mathrm{e}},K_{\mathrm{l}})=(2,8,32,7)$, corresponding to $K_{\mathrm{e}}=N$, $K_{\mathrm{p}}=M$, and $K_{\mathrm{l}}=M-1$. We refer to this as the \emph{two-pass full search}: all receive points, RIS elements, and nearest-neighbor comparisons are included in each pass, although the final phase shift matrix remains a greedy solution rather than the global optimum of \eqref{eq:GMD_problem}.

\subsection{BER Performance Comparison}
\begin{figure}[t]
\centering
\includegraphics[width=\paperfigwidth]{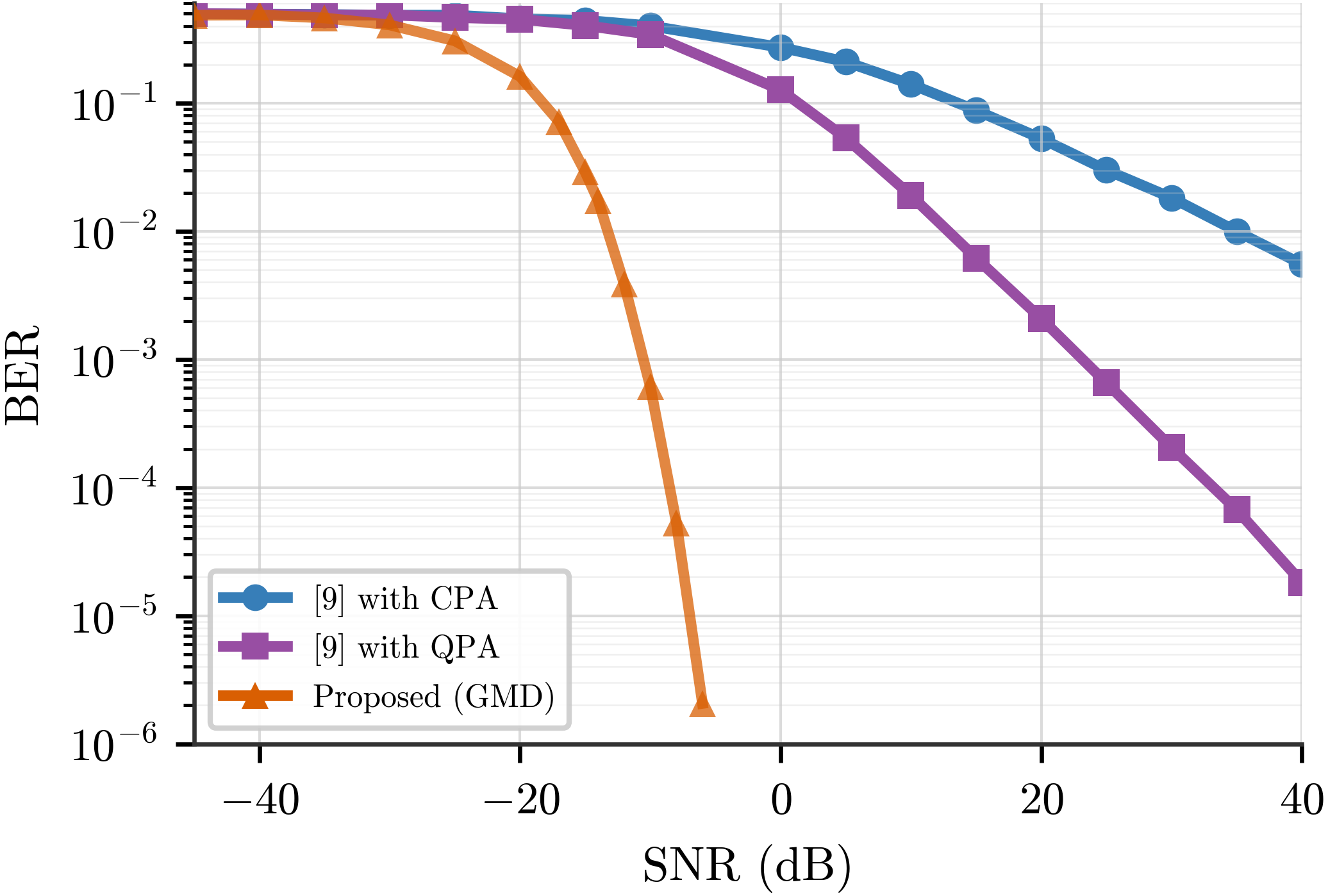}
\caption{Comparison of the BER of the proposed method with that of the method of~\cite{Yue2023TFSSK} with CPA and QPA. System parameters: $N_t=4$, $M=8$, $N=32$, $L=4$.}\label{fig:main_ber}
\vspace{-1mm}
\end{figure}

Fig.~\ref{fig:main_ber} compares the BER of the proposed optimization method with that of the method of~\cite{Yue2023TFSSK} with CPA and with QPA. At a target BER of $10^{-4}$, the proposed method exhibits a gain of \SI{84}{dB} and \SI{41.7}{dB} over the method of ~\cite{Yue2023TFSSK} with CPA and QPA, respectively. The BER curves of the proposed method also show a steeper asymptotic slope than the method of~\cite{Yue2023TFSSK} with QPA, indicating the larger minimum Euclidean distance in the designed receive signal constellation. We observe that the method of~\cite{Yue2023TFSSK} with QPA outperforms the corresponding method with CPA despite using only $L=4$ discrete phase shifts, which shows that gain-oriented alignment is not a suitable design objective for the case of ML-detected RIS-assisted SM once discrete phase shift control is enforced. The substantial gain relative to the method of~\cite{Yue2023TFSSK} with QPA confirms that, given the same discrete hardware constraint, switching from a gain-oriented to a distance-oriented objective is the main reason for improvement.

\subsection{Effect of the Number of RIS Elements Tested per Receive Point ($K_{\mathrm{e}}$)}
\begin{figure}[t]
\centering
\includegraphics[width=\paperfigwidth]{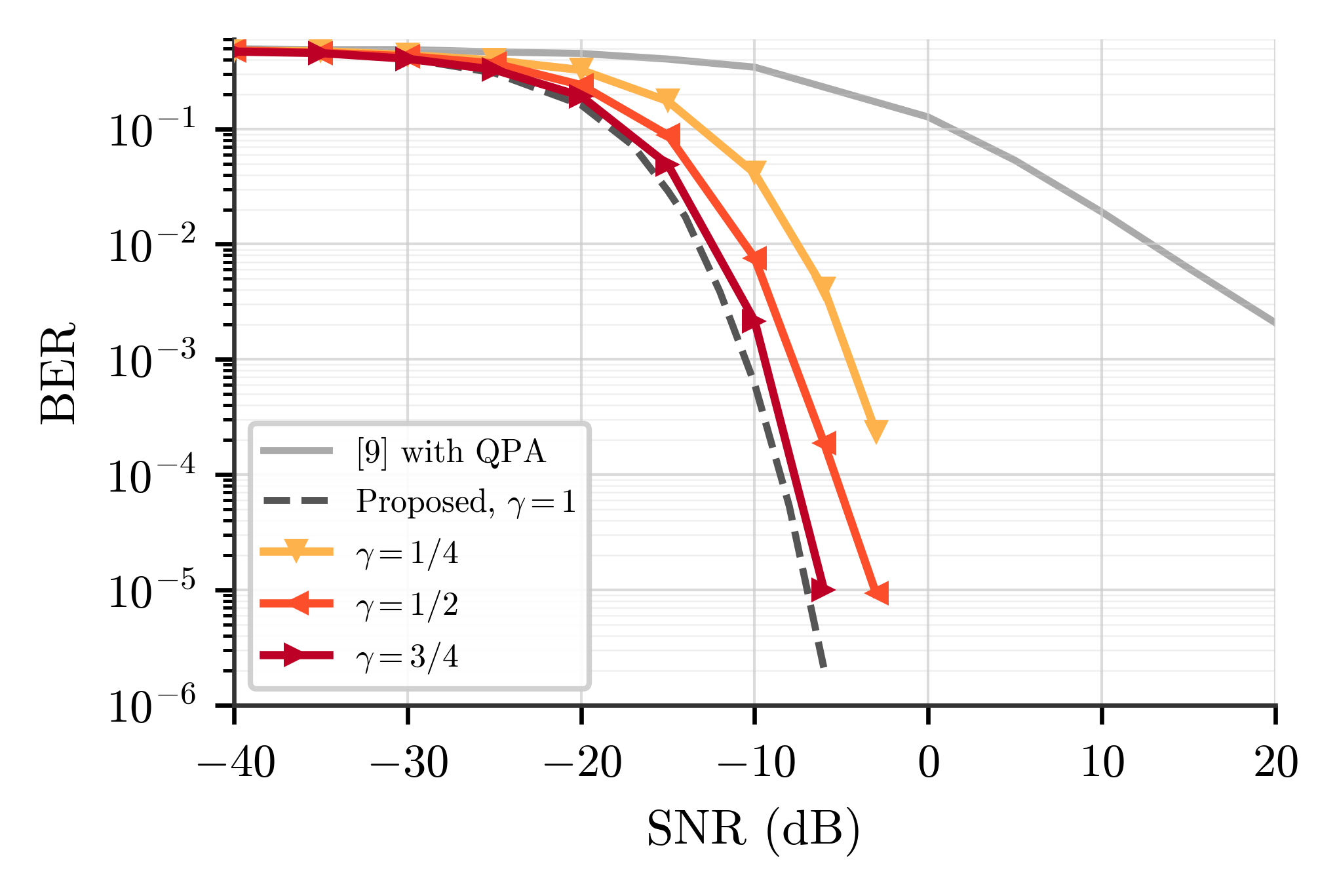}
\caption{Effect of the number of RIS elements tested per prioritized receive point, $K_{\mathrm{e}}$, on the BER performance of the proposed RIS-assisted SM system. Configuration: $(P,K_{\mathrm{p}},K_{\mathrm{l}})=(2,8,7)$ with variable $K_{\mathrm{e}}$.}\label{fig:kn_sweep}
\vspace{-1mm}
\end{figure}

The parameter $K_{\mathrm{e}}$ in Algorithm~\ref{alg:gmd_main} determines how many RIS elements are included in the search at each prioritized receive point, with priority given to the elements with the largest cascaded channel gains $\alpha_{m,n}$ as defined in Section~\ref{sec:system}. Define the relative budget $\gamma = K_{\mathrm{e}}/N \in (0,1]$. As shown in Fig.~\ref{fig:kn_sweep}, when $\gamma$ is reduced from the maximum-search value $\gamma = 1$ to $\gamma = 1/4$, the BER curves shift toward higher SNR. For small $\gamma$, the algorithm has access to too few elements to substantially shift the nearest-neighbor distance of the prioritized receive points, and the BER degrades noticeably. As $\gamma$ increases, the performance approaches the reference of $(P,K_{\mathrm{p}},K_{\mathrm{e}},K_{\mathrm{l}})=(2,8,32,7)$; in this sweep, it can be observed that $\gamma = 1/8$ requires approximately \SI{9.5}{dB} more SNR than the configuration $\gamma = 1$ to reach a BER of $10^{-4}$. Similar trends were observed for the parameters $K_{\mathrm{p}}$ and $K_{\mathrm{l}}$ (results are omitted due to space limitations).

\subsection{Comparison Under a Computational Complexity Constraint}
A natural concern is whether the BER advantage of the proposed method arises simply from its having higher computational complexity than the method of~\cite{Yue2023TFSSK} with QPA. To separate the effect of the design rule from that of computational budget, we restrict $(P,K_{\mathrm{p}},K_{\mathrm{e}},K_{\mathrm{l}})$ so that the dominant per-channel-realization cost of the proposed method does not exceed that of the quantized phase-alignment benchmark.

The computational complexity of the RIS phase shift optimization method of~\cite{Yue2023TFSSK} with QPA is given by $C_{\mathrm{QPA}} = 2 M_{\mathrm{tot}}N$, where the factor of 2 accounts for two length-$N$ passes in the algorithm: one pass to compute the cascaded channel terms and one pass to quantize/apply the phase shifts. 
The computational complexity of the proposed method is

\[
C_{\mathrm{prop}} \approx P\left[ M_{\mathrm{tot}}^2 (1 + K_{\mathrm{p}}) + 3 L K_{\mathrm{p}} K_{\mathrm{e}} K_{\mathrm{l}} \right].
\]

The term \(M_{\mathrm{tot}}^2(1+K_{\mathrm{p}})\) comes from comparing the \(M_{\mathrm{tot}}\) received points and repeating the distance check for the \(K_{\mathrm{p}}\) selected points. The term \(3LK_{\mathrm{p}}K_{\mathrm{e}}K_{\mathrm{l}}\) comes from testing \(L\) possible phase states for each of the \(K_{\mathrm{p}}K_{\mathrm{e}}\) point--element pairs against \(K_{\mathrm{l}}\) neighbors, with the factor of 3 representing the small set of arithmetic operations required for each trial distance evaluation. 

The complexity-matched configurations are therefore selected to satisfy $C_{\mathrm{prop}} \leq C_{\mathrm{QPA}}$. For the simulation setting \(M_{\mathrm{tot}}=8\), \(N=32\), and \(L=4\), this gives \(C_{\mathrm{QPA}}=2\cdot 8\cdot 32=512\), so the retained configurations are those whose computational complexity does not exceed 512 operations.

The constraint defines a set of feasible configurations rather than a single operating point. The retained feasible configurations
$(P,K_{\mathrm{p}},K_{\mathrm{e}},K_{\mathrm{l}})$ are
\begin{equation*}
\begin{gathered}
(1,1,8,4),\ (1,2,13,1),\ (1,3,7,1),\ (1,4,2,2),\\ 
(1,5,2,1),\ (2,1,5,2),\ (2,2,2,1),\ (3,1,3,1).
\end{gathered}
\end{equation*}

\begin{figure}[H]
\centering
\includegraphics[width=\paperfigwidth]{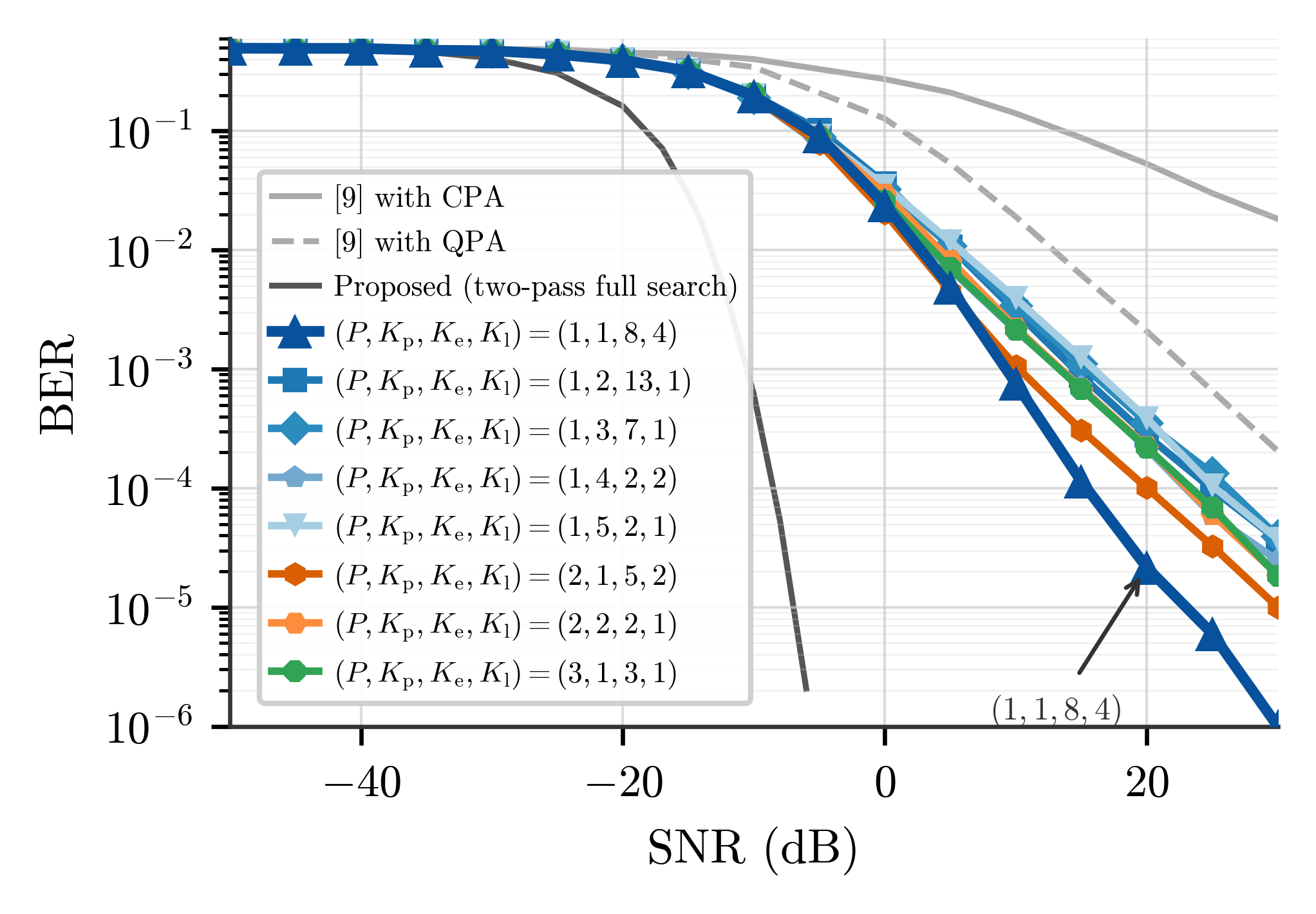}
\caption{Comparison of the BER performance of constrained-complexity configurations of the proposed method alongside the method of~\cite{Yue2023TFSSK} with CPA, the method of~\cite{Yue2023TFSSK} with QPA, and the reference search.}\label{fig:complexity_matched}
\vspace{-1mm}
\end{figure}

Fig.~\ref{fig:complexity_matched} compares the BER for these configurations with that of the method of~\cite{Yue2023TFSSK} with CPA and with QPA. For completeness we also show the performance of the proposed method with two-pass full search. We observe that each of these complexity-constrained implementations of the proposed method outperforms both phase-alignment baselines. Among these, $(1,1,8,4)$ achieves the lowest BER, reaching approximately $10^{-6}$ at \SI{10}{dB}. The pattern across configurations suggests that, under the complexity constraint, it is generally preferable to keep \(K_{\mathrm{p}}\) small and assign the remaining search effort to \(K_{\mathrm{e}}\) and \(K_{\mathrm{l}}\). We conclude that the improvement over the method of~\cite{Yue2023TFSSK} with QPA cannot be attributed to higher complexity; it persists even when the proposed method is restricted to the same operation count.

\begin{figure}[H]
\centering
\includegraphics[width=0.85\columnwidth]{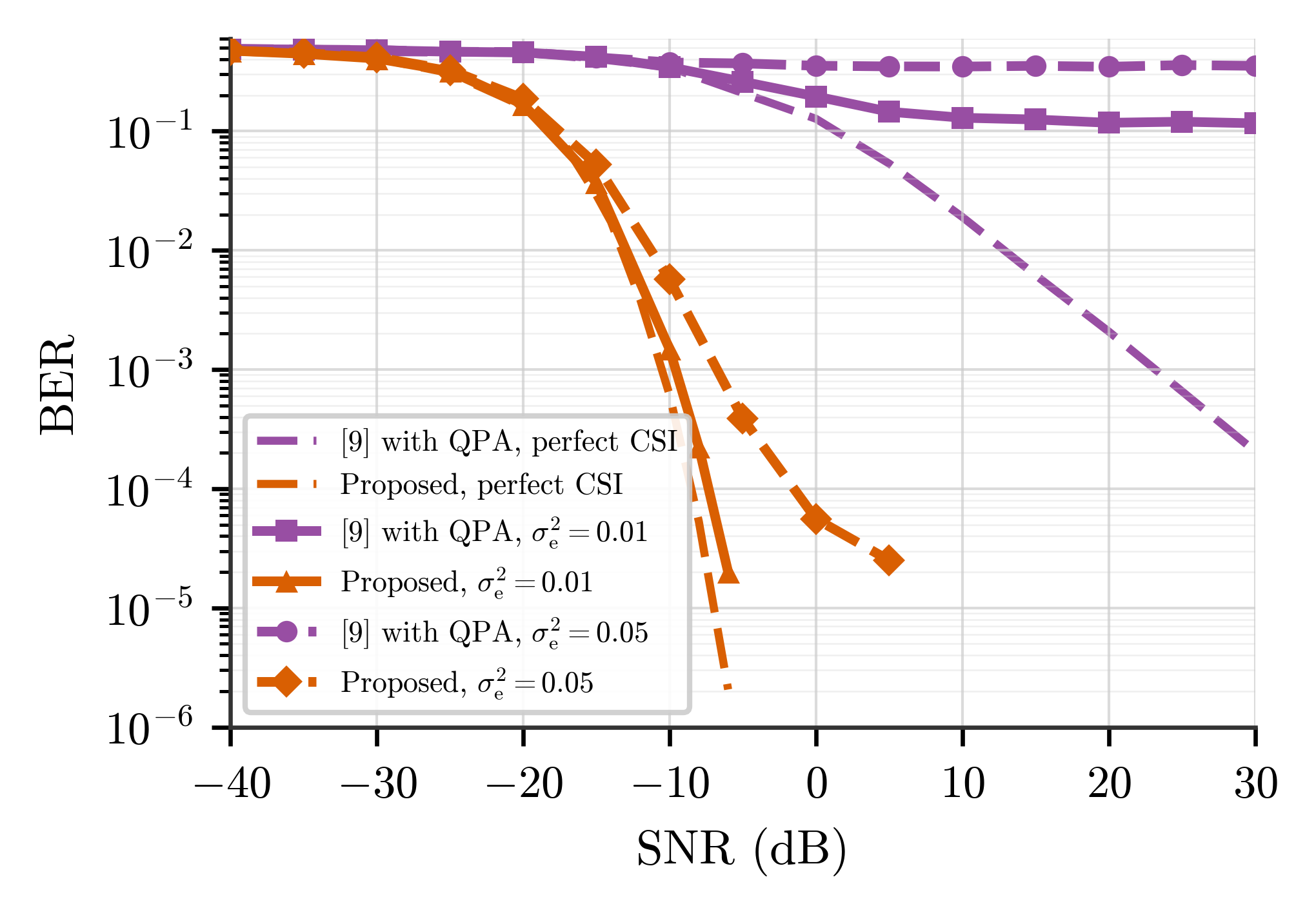}
\caption{BER performance comparison of the proposed method and the baselines under perfect and imperfect CSI for $\sigma_{\mathrm{e}}^2 \in \{0.01,0.05\}$.}\label{fig:imperfect}
\vspace{-1mm}
\end{figure}

\subsection{Robustness Under Imperfect CSI}\label{sec:imperfect}
Next, we evaluate the robustness of the proposed method under imperfect CSI. The channel estimates are modeled as $\hat{\mat{H}} = \mat{H} + \mat{E}$ and $\hat{\vect{g}} = \vect{g} + \vect{e}$, where each element of $\mat{E}$ and $\vect{e}$ is distributed according to $\mathcal{CN}(0,\sigma_{\mathrm{e}}^2)$. The RIS phase shift matrix is designed from the estimated channels, while transmission occurs over the true channels; the receiver applies ML detection against the receive-point set computed from the same estimates.

Fig.~\ref{fig:imperfect} shows the BER under imperfect CSI for $\sigma_{\mathrm{e}}^2 \in \{0.01,0.05\}$, with the case of perfect CSI as a benchmark. As shown in the figure, all schemes degrade relative to the perfect-CSI case, and BER plateaus appear at high SNR. In the latter regime, the remaining errors arise primarily from the geometric mismatch between the true receive-point set and the estimated set used by the detector, rather than from additive noise. Nevertheless, the proposed method continues to reach a lower BER than the method of~\cite{Yue2023TFSSK} with QPA across the SNR range examined, indicating that a phase shift configuration designed to maximize $d_{\min}$ using the estimated channels still tends to produce a more separated receive-point set under the true channels.

\vspace{-1mm}
\section{Conclusion}\label{sec:conclusion}
In this paper, we have proposed a reduced greedy search algorithm for RIS phase shift design in RIS-assisted SM systems with discrete phase shift control. The proposed method targets the minimum Euclidean distance of the ML receive-point set rather than the effective channel gain. Numerical results show that the proposed method achieves a significant improvement in the BER performance relative to the method of~\cite{Yue2023TFSSK} with QPA under the same discrete hardware model, and that this gain persists even when the proposed method is restricted to the same receiver algorithm complexity. The proposed method also continues to outperform the gain-oriented baselines under imperfect CSI. Extensions to multi-antenna receivers, frequency-selective channels, and joint design with channel estimation are natural directions for further investigation.

\section{Acknowledgment}
This research was supported by Taighde Éireann -- Research Ireland under Grant Number 24/FFP-P/12895.

\end{document}